# Space-Time Scattering Network for Electromagnetic Inverse Design and Tomography

*Travis Hamilton and Hooman Mohseni*

*Department of Electrical Engineering, Northwestern University, Evanston , IL*

**Maxwell's equations generally explain the propagation of light through an arbitrary medium by using wave mechanics. However, scientific evidence since Newton's time suggest a discrete interpretation of light more generally explains its nature. This interpretation lends itself well to the discrete form of computer simulation. While current simulations attempt to discretize Maxwell's equations, we present an inherently discrete physical model of light propagation that naturally forms a causal space-time scattering network (STSN). Since STSN has the topology of neural networks, inverse design and tomography based on STSN can be readily implemented in a variety of software and hardware that are optimized for deep learning. Also, STSN inherently includes the physics of light propagation, and hence the number of unknown weights in STSN is at a minimum. We show this property leads to orders of magnitude smaller number of unknown weights, and a much faster convergence, compared with inverse design methods using conventional neural networks. In addition, the intrinsic presence of space-time fabric in STSN allows time-dependent inverse design and tomography. We show examples of the fast convergence of STSN in predicting time-dependent index profiles while avoiding approximations typically used.**

Light propagating through a vacuum has motivated scientific research for over three centuries. This baffling phenomenon was explained through the Luminiferous aether theory, or the concept of an ever-present but undetectable medium that facilitated light propagation in the absence of matter. Aether lacks empirical evidence [1] and was abandoned for special relativities simple axiom that the speed of light in vacuum is constant. The particle-based theory of light allows for particle-particle interactions, which naturally give way to discrete models and relies on a sea of background particles to facilitate light propagation. Such a view of light propagation is in harmony with modern physics [2] [3] [4]. However, Maxwell's equations, and the continuous wave equation it implies, are the most ubiquitous models of light propagation. Unfortunately, the wave equation does not provide an inherently discrete physical model, and forces computer simulations to approximate discrete forms of Maxwell's equation [5] [6] [7] [8] [9].

Here we utilize the non-wave view of light propagation to construct an inherently discrete and highly efficient computational model to compute light propagation through a medium (including vacuum) with arbitrary space and time-dependent optical properties (i.e. the forward model), and to predict a possible medium that can generate a given optical response (i.e. inverse model). Due to its predictive nature, the model is highly relevant in fields requiring inverse design to predict unknown structures and create non-intuitive devices.

Inverse design has emerged as an exciting methodology for creating highly complex and compact photonic and microwave devices [10] [11] [12] [13]. The goal of inverse design, broadly speaking, is to solve for a suitable material design that approximates a desired electromagnetic response. Tomography [14] [15] [16] is similar to inverse design in nature, except that a unique solution is sought after.

The existing tomography and inverse design methods belong to three categories: physics-based optimization, physics-based optimization using neural networks, and non-physics optimization using neural networks. In physics-based optimization [11] [17] [18], the input/output characteristics are optimized to match a desired response by adjusting the refractive index of space while using Maxwell's equations as a constraint. While highly versatile, the method is not efficient for time-dependent problems and optimizes for one frequency at a time. In physics-based optimizations a neural network is used to implement a physical theory of light propagation. The network's weights are then trained by comparing the network's forward propagating output with the desired output. However, all existing methods use limiting approximations such as beam propagation approximation [14], Born approximation [15], and the Lippmann-Schwinger approximation [16]. Therefore, they cannot be used for many important conditions, such as high index contrast and a strong backscattering. Non-physics optimization using neural networks [13] [19] [20] [21] define a topological parameter of interest, train a neural network given electromagnetic inputs and outputs to learn the physics of the topology for a range of parameter values and optical constants, and then invert the network to produce a topology that matches a desired input/output characteristic. Once a network is trained it can produce a design much faster than the previously mentioned methods. However, this approach is extremely inefficient computationally, and requires large data sets to train properly.

Here we present a novel approach we call space-time scattering network (STSN) that is based on the general scattering model of light propagation, and without any simplification. We show that STSN can be efficiently implemented by commonly used software (open source TensorFlow), and hardware (CPU/GPU) for forward modeling, inverse design, and tomography in broadband and time-dependent optical systems that include large index contrasts and strong backscattering.

We start by assuming light scattering at subwavelength points within the media, including vacuum. For simplicity, here we assume the scattering points are equally spaced on a Cartesian grid. Note that neither assumptions are necessary in general. Assuming causality holds, we now arrange these scattering events across time, producing a "static" space-time fabric. We start by developing STSN in one-dimensional space and show how to expand it to higher dimensional space mathematically (we limit our graphical representation to 1D and 2D space, since the 3D space-time graph is not visually representable).

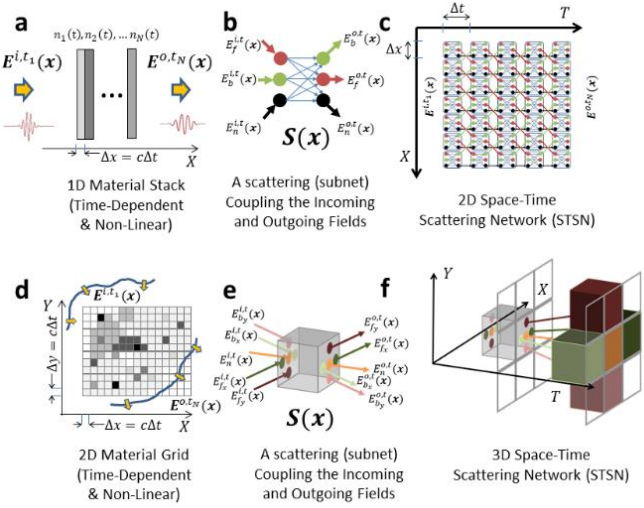

Figure 1: The STSN model. STSN in 1D solves for the refractive index of a stack of materials (a) using the final outgoing and initial incoming fields. The scattering element of the 1D STSN (b) connects the field components of the outgoing and incoming fields at each point in space and is connected through the translation operator to form the full network (c). Similar, in 2D a grid of refractive index values are solved for (d) using a scattering subnet (e) that now takes into account the x and y directions. The full network is shown in (f).

Figure 1a shows the essence of the 1D STSN model: a stack of sub-wavelength thick material slabs extending infinitely in the direction perpendicular to propagation and with their own unique refractive index values to be solved using incoming and outgoing fields. The scattering of light is represented mathematically by a tensor field $S(x)$ at a location given by the position vector $x$. It couples the incoming and outgoing fields which are compactly written as tensor fields $E^{i,t}(x)$ and $E^{o,t}(x)$ and have field components forward incoming/outgoing $E_f^{i/o,t}(x)$, backward incoming/outgoing $E_b^{i/o,t}(x)$ and node (local) incoming/outgoing $E_n^{i/o,t}(x)$ (Figure 1b). The forward and backward directions are with respect to the positive $X$ direction in Figure 1a. The tensor field $S(x)$ at a given location is represented as a subnet and will become the fundamental building block of the STSN. The connections of scattering subnets are dictated by a translation operator T and is shown visually in Figure 1c.

The 2D STSN follows from the 1D STSN. Now a gird of unknown refractive index materials is solved (Figure 1d) and the scattering subnet (Figure 1e) couples the $X$ and $Y$ field components. The network itself is 3D (Figure 1f) due to the two dimensions needed for space and one dimension needed for time. As in the 1D case, a translation operator defines the connections between scattering subnets.

Mathematically, the STSN can be described using the well understood transmission line matrix theory (TLM) developed by Peter Johns [5]. The STSN model can be expressed in a compact form for one, two and higher dimensions. The scattering subnet is represented as a tensor field coupling the incoming and outgoing tensor fields: $E^{o,t}(x) = S(x)E^{i,t}(x)$. The connection to the next time step can be defined via a translation operator T, such that: $E^{i,t+\Delta t}(x) = T(E^{o,t}(x))$. Note that tensor field $S$ is only related to the optical properties of the material, while operation $T$ is simply producing interconnection between geometrically adjacent scattering points in space (details of S and T are presented in the Supplementary Material).

To evaluate the accuracy of STSN model, we compare simulation results of 1D and 2D examples with results produced by finite difference time domain (FDTD Solutions, Lumerical Inc.). Identical structures were simulated in both STSN and Lumerical and the resulting fields show good agreement (Figure 2). The overlap integral value between the Lumerical and STSN electric fields across all simulated space are 98.74% and 95.41% for the 1D and 2D cases respectively.

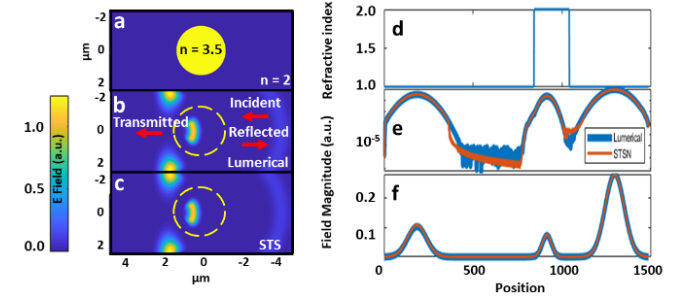

Figure 2: STSN validated by Lumerical software package. (a) shows the refractive index distribution for the 2D test case. (b) and (c) show the 2D test case for Lumerical and STSN. (d) shows the refractive index distribution for the 1D test case. (e) and (f) show the 1D test case for Lumerical.

STSN naturally forms a network that has a forward flow in the direction of the "arrow of time," and can be readily implemented by any software/hardware designed for neural networks (NN) and deep learning. Note that the number of unknown weights in the resulting neural network is exactly equal to the number of unknown optical parameters in the system (i.e. number of parameters in the design space), which is the minimum possible number of unknown weights. This is in stark contrast with the non-physics based neural networks, where the number of unknown weights is typically much larger than the number of parameters in the design space, since many of the weights are used by the network to "learn" the physics involved.

The STSN is implemented as a NN in Google's open source TensorFlow where the graph's layers are defined by scattering subnet $S(x)$ and translation operator T. The scattering tensor field at one point in space $S(x = (x_i, y_j, z_k))$ produces a two-layer subnet (Figure 1-b and e). This subnet represents the optical properties of one point in space and its weights are directly related to the refractive index (See Supplementary Material). The subnets are linked through $T$, which pushes the network through time (Figure 1-c and f). Note that the subnets weights may differ across time as the refractive index changes with time. The weights are initialized to free space. An area of interest in the graph is defined across space where the refractive index is unknown and trainable. All weights outside the region are left constant.

To evaluate the tomography and inverse design capabilities of STSN, both static and time-dependent refractive index distributions are trained, using the optical response of the media to incident pulses of light (Figure 3a-d and f-i). The percent error average and standard deviation between the trained and ground truth refractive index distributions is calculated to quantify the prediction error (i.e. a figure of merit for tomography) and are shown in Table 1. Note that for time-dependent refractive indices the model cannot reconstruct their values for the final five time steps and when they are excluded, an average percent error below 1% is achieved for all examples. The error between

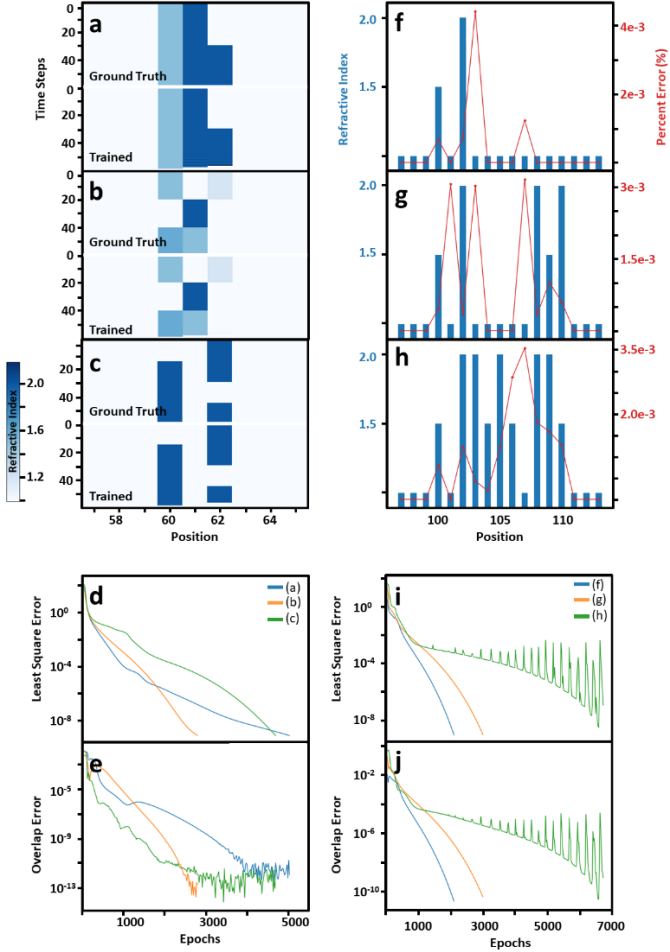

| Figure 3 | Unknowns | Time Changes | Average of Percent Error | Standard Deviation of Percent Error |
|---|---|---|---|---|
| a | 180 | 1 | 3.73% | 18.27% |
| b | 180 | 2 | 0.92% | 6.85% |
| c | 180 | 3 | 2.23% | 14.74% |
| f | 11 | 0 | 0.00064% | 0.0013% |
| g | 11 | 0 | 0.0011% | 0.0013% |
| h | 11 | 0 | 0.0014% | 0.0010% |
| a* | 180 | 1 | 0.64% | 7.76% |
| b* | 180 | 2 | 0.0084% | 0.048% |
| c* | 180 | 3 | 0.0023% | 0.0089% |

Table 1: The percent error average and standard deviation between the ground truth and trained refractive index distributions shown in figure 3. *Excluding the 5 final time steps.

An inversely designed notch filter example was taken from one of the best reported non-physics based optimizations using neural networks [21] to demonstrate the performance of STSN (Figure 4). Our model accurately produces the notch filter while using four orders of magnitude less weights and one training set (see Table 2).

| Model | Layers | Weights | Training Sets | Saturation Epoch |
|---|---|---|---|---|
| STSN | 200 | 60 | 1 | 100 |
| Ref. [21] | 6 | >300,000 | 550,000 | 2000 |

Table 2: Comparison of STSN's and NPNN's computing resources needed to inversely design a notch filer.

To ensure an accurate comparison, the STSN model was constrained to 3ums with 50nm resolution which is comparable to the 1.1ums and 8nm resolution used in [21]. An identical transmittance spectrum was used with fewer frequency points (20) and the structure was set to represent vacuum initially.

We attribute the many orders of magnitude lower computational cost of STSN in this inverse design example to its physics-based topology. The single training set suggests a causal network is achieved while the smaller weight count exhibits the efficiency of the STSN design. A physics-based approach does not waste resources on learning the physics of light propagation. For tomography applications, STSN is shown to accurately recreate features densely packed in both time and space even with high refractive index profiles. While previous tomography models must resort to approximations with small backscattering, STSN appears capable of handing optically complex structures that are also time-dependent.

Figure 3: Results for time-dependent and time-independent refractive index distributions. Time-dependent results for (a) 3 scatterers and 1 time change (b) 3 scatterers and 2 time changes and (c) 3 scatterers and 3 time changes. In all three cases, there are 180 unknowns and the ground truth distribution is shown above the trained distribution. Time independent results are shown for (f) 2 scatterers (g) 5 scatterers and (h) 9 scatterers. In all three cases there are 11 unknowns and the percent error between trained and ground truth for each position is shown in red. The region of trainable refractive index values range from 60 to 62 for the time dependent case and 100 to 110 for the static case. The least squares (L.S.) between the ground truth and trained refractive index values are shown. The least squares error between ground truth and trained output fields for time-dependent (d) and time-independent (i) are shown along with the overlap integral over all time at the point directly after the material for time-dependent (e) and time-independent (j) results.

the desired and achieved optical response was also quantified as the sum of the difference in space-time (overlap integral) between the desired and achieved responses at the point in space directly after the material, to quantify the inverse design error (Figure 3e,j). The training set is a single field over all space at the final time step. For the static case, only one training field is needed. For the time-dependent case, the field's initial position is swept to produce a staggered set of fields at the final time step. The training sets are produced by running the STSN forward in time with the desired weight distribution.

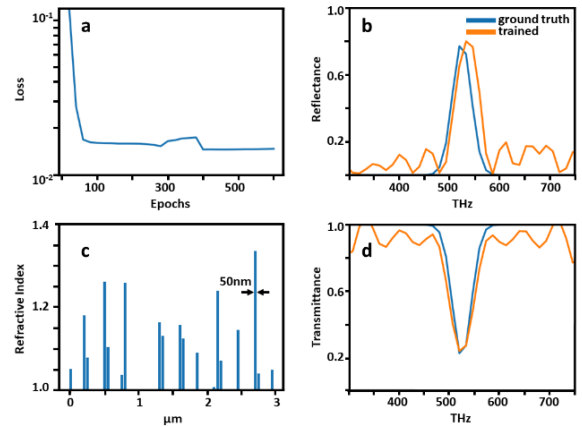

Figure 4: A notch filter designed using STSN. (a) The cost function between the transmitted and expected spectrums. (b) and (d) show the resulting reflectance and transmittance of the filter while (c) shows the resulting dielectric structure with 50nm resolution.

This paper introduces space-time scattering network STSN for inverse design and tomography. STSN can be used for inverse design and tomography of structures that are time-dependent and anisotropic. The network is implement in an open-source software (Tensorflow), and can take advantage of the state-of-the-art hardware, such as GPUs. We show that STSN can reproduce several refractive index structures that were both static and time-dependent. STSN also was compared with one of the best reported inverse design methods based on neural network. STSN achieves the inverse design with orders of magnitude fewer weights and a much faster convergence rate. While our examples are based on 1D structures, STSN is easily translated into higher dimensions, as shown in the Supplementary Material. In addition, we showed that optical polarization and anisotropy of the material can also be included in 3D STSN seamlessly.

**Funding.** Army Research Office award number W911NF11-1-0390.

**Acknowledgment**. We thank Professor Katsaggelos for his help with the Tensorflow implementation of STSN.

## Supplementary Materials

$E^{o,t}$ **and** $E^{i,t}$ **Tensors:** $E^{o,t}(x)$ and $E^{i,t}(x)$ are tensor fields of rank 1 with position vector $x$ at time $t$. For 1D and 2D, the tensor contains the field components at each point in space and time. The field components are forward $E_f$ and backward $E_b$ for each Cartesian direction (x and y) as well as internal node components $E_n$.

$$E^{o/i,t}(x) = [E_{fx}^{o/i,t}(x), E_{fy}^{o/i,t}(x), E_{bx}^{o/i,t}(x), E_{by}^{o/i,t}(x), E_n^{o/i,t}(x)]$$

In the third dimension, the internal field is dropped, and polarization is introduced. Therefore, there are now four field components for each direction; two for direction and two for polarization.

**The $S$ Tensor and T Operation:** $S(x)$ is a tensor field of rank 2 that transforms the input and output fields at each point in space through a scattering matrix.

$$E^{o,t}(x) = S(x) E^{i,t}(x)$$

For 1D, $S(x)$ is equal to [5]

$$S(x = x) = \frac{1}{2 + W(x)} \begin{bmatrix} 2 & -W(x) & 2W(x) \\ -W(x) & 2 & 2W(x) \\ 2 & 2 & W(x) - 2 \end{bmatrix}$$

Where $W(x)$ is a scalar field and represents the weight values in space. This weight is trained by the neural network and is related to the physical dielectric constant through the following equation [5]

$$\epsilon_r(x) = 1 + \frac{W(x)}{2}$$

For 2D, $S(x)$ is shown in (S1) [5], where $W(x, y)$ is a scalar field and represents the weight values in space. This weight is related to the physical dielectric constant through the following equation [5]

$$\epsilon_r(x, y) = 2 + \frac{W(x, y)}{2}$$

Three dimensions can be easily constructed using the transmission line matrix (TLM) method's symmetrical super-condensed node [22]. The node introduces polarizations and eliminates the need for internal node fields. More recent work has focused on dispersive mediums and could be implemented in STSN method [3]. The scatter tensor field is **Error! Reference source not found.**. Subscripts have been used in place of standard functional notation for readability and are defined as,

$$\begin{aligned} a_{ij} &= 1 - b_{ij} - d_{ij} \\ c_{ij} &= d_{ij} - b_{ij} \\ b_{ij} &= \hat{C}_{kj} \\ d_{ij} &= \hat{C}_{ik} \end{aligned}$$

Where $\hat{C}$ is a scalar field and is related to the dielectric constant with polarization direction $j$ by

$$\hat{C}_{ik} \hat{C}_{ij} = \left(\frac{\Delta t}{\Delta i}\right)^2 \frac{c^2}{\varepsilon_j}$$

Where $\Delta t$ is the time step, $\Delta i$ the spatial step in direction $i$, $\varepsilon_j$ the relative dielectric constant in polarization direction $j$, and $c$ is the speed of light in vacuum. Note that anisotropic materials are allowed in both permeability and permittivity. However, here we assume anisotropic permittivity only due to the simplicity of its relationship to the scattering matrix's weights.

The Translation operator is defined as follows

$$E_j^{i,t+\Delta t}(x + \Delta x) = E_j^{o,t}(x + \Delta x) = T_j E_j^{o,t}(x)$$

Where $j$ is the element of the rank 1 tensor field $E$ making $T_j$ a two tensor or matrix.

$$S(x = [x,y]) = \frac{1}{4 + W(x,y)} \begin{bmatrix} 2 & 2 & -W(x,y)-2 & 2 & 2W(x,y) \\ 2 & 2 & 2 & -W(x,y)-2 & 2W(x,y) \\ -W(x,y)-2 & 2 & 2 & 2 & 2W(x,y) \\ 2 & -W(x,y)-2 & 2 & 2 & 2W(x,y) \\ 2 & 2 & 2 & 2 & W(x,y)-4 \end{bmatrix}$$

(S1)

$$S(x = [x,y,z]) = \begin{bmatrix} a_{xy} & b_{xy} & d_{xy} & - & - & - & - & b_{xy} & - & -d_{xy} & c_{xy} \\ b_{xz} & a_{xz} & - & - & - & d_{xz} & - & - & c_{xz} & -d_{xz} & - & b_{xz} \\ d_{yx} & - & a_{yx} & b_{yx} & - & - & - & b_{yx} & - & - & c_{yx} & -d_{yx} \\ - & - & b_{yz} & a_{yz} & d_{yz} & - & -d_{yz} & c_{yz} & - & - & b_{yz} & - \\ - & - & - & d_{zy} & a_{zy} & b_{zy} & c_{zy} & -d_{zy} & - & b_{zy} & - & - \\ - & d_{zx} & - & - & b_{zx} & a_{zx} & b_{zx} & - & -d_{zx} & c_{zx} & - & - \\ - & - & - & -d_{zy} & c_{zy} & b_{zy} & a_{zy} & d_{zy} & - & b_{zy} & - & - \\ - & - & b_{yz} & c_{yz} & -d_{yz} & - & d_{yz} & a_{yz} & - & - & b_{yz} & - \\ b_{xz} & c_{xz} & - & - & - & -d_{xz} & - & - & a_{xz} & d_{xz} & - & b_{xz} \\ - & -d_{zx} & - & - & b_{zx} & c_{zx} & b_{zx} & - & d_{zx} & a_{zx} & - & - \\ -d_{yx} & - & c_{yx} & b_{yx} & - & - & - & b_{yx} & - & - & a_{yx} & d_{yx} \\ c_{xy} & b_{xy} & -d_{xy} & - & - & - & - & b_{xy} & - & d_{xy} & a_{xy} \end{bmatrix}$$

(S2)